\documentclass[prl,twocolumn]{revtex4}

\usepackage{amsmath}
\usepackage[english]{babel}

\def\ket#1{{|#1\rangle}}

\def\tr{{\rm Tr}}

\begin{document}

\title{Weak measurements are universal}
\author{Ognyan Oreshkov}
\email{oreshkov@usc.edu}
\affiliation{Department of Physics, University of Southern California, \\
Los Angeles, CA  90089}

\author{Todd A. Brun}
\email{tbrun@usc.edu}
\affiliation{Communication Sciences Institute, University of Southern California, \\
Los Angeles, CA  90089}

\date{\today}

\begin{abstract}
It is well known that any projective measurement can be decomposed
into a sequence of {\it weak} measurements, which cause only small
changes to the state.  Similar constructions for generalized
measurements, however, have relied on the use of an ancilla
system.  We show that any generalized measurement can be
decomposed into a sequence of weak measurements without the use of
an ancilla, and give an explicit construction for these weak
measurements.  The measurement procedure has the structure of a
random walk along a curve in state space, with the measurement
ending when one of the end points is reached.  This shows that any
measurement can be generated by weak measurements, and hence that
weak measurements are universal.  This may have important
applications to the theory of entanglement.
\end{abstract}

\maketitle


In the original formulation of measurement in quantum mechanics,
measurement outcomes are identified with a set of orthogonal
projection operators, which can be thought of as corresponding to
the eigenspaces of a Hermitian operator, or {\it observable}
\cite{vonNeumann,Lueders}.  After a measurement, the state is
projected into one of the subspaces with a probability given by
the square of the amplitude of the state component in that
subspace.

In recent years a more general notion of measurement has become
common:  the {\it generalized} or {\it positive-operator valued}
measurement (POVM) \cite{POVM}.  This formulation can include many
phenomena not captured by projective measurements:  detectors with
non-unit efficiency, measurement outcomes that include additional
randomness, measurements that give incomplete information, and
many others.  POVMs have found numerous applications, especially
in the rapidly-growing field of quantum information processing
\cite{NielsenChuang00}.

Upon measurement, a system with density matrix $\rho$ undergoes a
random transformation
\begin{equation}
\rho \rightarrow \rho_j= \hat{M}_j\rho
\hat{M}_j^{\dagger}/p_j, \hspace{0.2cm}
\underset{j}{\sum}\hat{M}_j^{\dagger}\hat{M}_j=\hat{I},
\end{equation}
with probability $p_j=\textrm{Tr}(\hat{M}_j \rho
\hat{M}_j^{\dagger})$, where the index $j$ labels the possible
outcomes of the measurement. This transformation is commonly
comprehended as a spontaneous {\it jump}, unlike unitary
transformations, for example, which are thought of as resulting
from {\it continuous} unitary evolutions. Any unitary transformation
can be implemented as a sequence of {\it weak}
(i.e., infinitesimal) unitary transformations. One may ask if a similar
decomposition exists for generalized measurements. This would
allow us to think of POVMs as resulting from continuous stochastic
evolutions and possibly make use of the powerful tools of
differential calculus in the study of the transformations that a
system undergoes upon measurement.

In this paper we show that any generalized measurement can be
implemented as a sequence of {\it weak measurements}
\cite{Weak,Brun02}. We call a measurement {\it weak} if all
outcomes result in very small changes to the state.  (There are
other definitions of weak measurements that include the
possibility of large changes to the state with low probability; we
will not be considering measurements of this type.) Therefore, a
weak measurement is one whose operators can be written as
\begin{equation}
\hat{M}_j=q_j(\hat{I}+\hat{\varepsilon}_j),
\end{equation}
where $0\leq q_j \leq 1$ and $\hat{\varepsilon}$ is an operator
with small norm $\|\hat{\varepsilon}\| \ll 1$.  Weak measurements
have been studied both in the abstract, and as a means of
understanding systems with continuous monitoring.  In the latter
case, we can think of the evolution as the limit of a
sequence of weak measurements, which gives rise to continuous
stochastic evolutions called {\it quantum trajectories}
\cite{QTraj,Brun02}.

It has been shown that any projective measurement can be done as a
sequence of weak measurements; and by using an additional {\it
ancilla} system and a joint unitary transformation, it is possible
to do any generalized measurement using weak measurements
\cite{Bennett99}. This procedure, however, does not decompose the
operation on the original system into weak operations, since it
uses operations acting on a larger Hilbert space---that of the
system plus the ancilla.  If we wish to study the behavior of  a
function---for instance, an entanglement monotone---defined on a
space of a particular dimension, it complicates matters to add and
remove ancillas.  We will show that an ancilla is not needed, and
give an explicit construction of the weak measurement operators
for any generalized measurement that we wish to decompose.

It is easy to show that a measurement with any number of outcomes
can be performed as a sequence of measurements with two outcomes.
Therefore, for simplicity, we will restrict our considerations to
two-outcome measurements.  To give the idea of the construction,
we first show how every projective
measurement can be implemented as a sequence of weak generalized
measurements. In this case the measurement operators $\hat{P}_1$
and $\hat{P}_2$ are orthogonal projectors whose sum
$\hat{P}_1+\hat{P}_2=\hat{I}$ is the identity. We introduce the
operators
\begin{equation}
\hat{P}(x)=\sqrt{\frac{1-\tanh(x)}{2}}\hat{P}_1+\sqrt{\frac{1+\tanh(x)}{2}}\hat{P}_2,
\hspace{0.5cm} x\in R.
\label{measpro}
\end{equation}
Note that $\hat{P}^2(x)+\hat{P}^2(-x)=\hat{I}$ and therefore
$\hat{P}(x)$ and $\hat{P}(-x)$ describe a measurement. If
$x=\varepsilon$, where $|\varepsilon| \ll 1$, the measurement is
weak. Consider the effect of the operators $\hat{P}(x)$ on a pure
state $|\psi\rangle$. The state can be written as
$|\psi\rangle=\hat{P}_1|\psi\rangle+\hat{P}_2|\psi\rangle=\sqrt{p_1}|\psi_1\rangle+\sqrt{p_2}|\psi_2\rangle$,
where
$|\psi_{1,2}\rangle=\hat{P}_{1,2}|\psi\rangle/\sqrt{p_{1,2}}$ are
the two possible outcomes of the projective measurement and
$p_{1,2}=\langle\psi|\hat{P}_{1,2}|\psi\rangle$ are the
corresponding probabilities. If $x$ is positive (negative), the
operator $\hat{P}(x)$ increases (decreases) the ratio
$\sqrt{p_2}/\sqrt{p_1}$ of the $|\psi_2\rangle$ and
$|\psi_1\rangle$ components of the state. By applying the same
operator $\hat{P}(\varepsilon)$ many times in a row for some fixed
$\varepsilon$, the ratio can be made arbitrarily large or small
depending on the sign of $\varepsilon$, and hence  the state can be transformed
arbitrarily close to $|\psi_1\rangle$ or $|\psi_2\rangle$. The
ratio of the $p_1$ and $p_2$ is the only parameter needed
to describe the state, since $p_1+p_2=1$.

Also note that $\hat{P}(-x)\hat{P}(x)=(1-\tanh^2(x))^{1/2}\hat{I}/2$
is proportional to the identity. If we apply
the same measurement $\hat{P}(\pm\varepsilon)$ twice and two opposite outcomes
occur, the system returns to its previous
state. Thus we see that the transformation of the state under many
repetitions of the measurement $\hat{P}(\pm\varepsilon)$ follows a random walk along a curve $\ket{\psi(x)}$ in state space.  The position on this curve can be parameterized by
$x=\ln\sqrt{p_1/p_2}$. Then $\ket{\psi(x)}$ can be written as
$\sqrt{p_1(x)}\ket{\psi_1} + \sqrt{p_2(x)}\ket{\psi_2}$, where
$p_{1,2}(x) = (1/2)[1 \pm \tanh(x)]$.

The measurement given by the operators $\hat{P}(\pm\varepsilon)$
changes $x$ by $x\rightarrow x\pm\varepsilon$, with probabilities
$p_{\pm}(x)=(1\pm\tanh(\varepsilon)(p_1(x)-p_2(x)))/2$.  We
continue this random walk until $|x| \ge X$, for some $X$ which is
sufficiently large that $\ket{\psi(X)} \approx \ket{\psi_1}$ and
$\ket{\psi(-X)} \approx \ket{\psi_2}$ to whatever precision we
desire.  What are the respective probabilities of these two
outcomes?

Define $p(x)$ to be the probability that the walk will end at $X$
(rather than $-X$) {\it given} that it began at $x$.  This must
satisfy $p(x) = p_+(x) p(x+\varepsilon) + p_-(x)
p(x-\varepsilon)$. Substituting our expressions for the
probabilities, this becomes
\begin{eqnarray}
p(x) &=& (p(x+\varepsilon) + p(x-\varepsilon))/2
\label{difference}
\\&&  +
\tanh(\varepsilon)\tanh(x)(p(x+\varepsilon)-p(x-\varepsilon))/2.\nonumber
\end{eqnarray}

If we go to the infinitesimal limit $\varepsilon\rightarrow dx$,
this becomes a continuous differential equation
\begin{equation}
\frac{d^2p}{dx^2} + 2\tanh(x)\frac{dp}{dx} = 0 ,
\end{equation}
with boundary conditions $p(X)=1$, $p(-X)=0$.  The solution to
this equation is $p(x)=(1/2)[1 + \tanh(x)/\tanh(X)]$. In the limit
where $X$ is large, $\tanh(X)\rightarrow1$, so $p(x)=p_1(x)$. The
probabilities of the outcomes for the sequence of weak
measurements are exactly the same as those for a single projective
measurement. Note that this is also true for a walk with a
step size that is not infinitesimal, since the solution $p(x)$ satisfies
\eqref{difference} for an arbitrarily large $\varepsilon$.

Alternatively, instead of looking at the state of the system
during the process, we could look at an operator that effectively
describes the system's transformation to the current state.
This has the advantage that it is state-independent,
and will lead the way to decompositions of generalized measurements;
it also becomes obvious that the procedure works for mixed states, too.

We think of the measurement
process as a random walk along a curve $\hat{P}(x)$ in operator
space, given by Eq.~(\ref{measpro}), which satisfies $\hat{P}(0)=\hat{I}/\sqrt{2}$,
$\underset{x\rightarrow -\infty}{\lim}\hat{P}(x)=\hat{P}_1$,
$\underset{x\rightarrow \infty}{\lim}\hat{P}(x)=\hat{P}_2$. It can
be verified that $\hat{P}(x)\hat{P}(y)\propto \hat{P}(x+y)$,
where the constant of proportionality is
$(\cosh(x+y)/2\cosh(x)\cosh(y))^{1/2}$. Due to normalization of
the state, operators which differ by an overall factor
are equivalent in their effects on the state.  Thus, the
random walk driven by weak measurement operators
$\hat{P}(\pm\varepsilon)$ has a step size $|\varepsilon|$.

Next we consider measurements where the measurement
operators $\hat{M}_1$ and $\hat{M}_2$ are {\it positive} but not projectors.
We use the well known fact that a generalized measurement
can be implemented as joint unitary operation on the system and an
ancilla, followed by a projective measurement on the ancilla
\cite{NielsenChuang00}.  (One can think of this as an {\it indirect}
measurement; one lets the system interact with the ancilla, and then
measures the ancilla.)
Later we will show that the ancilla is not needed.
We consider two-outcome measurements and two-level ancillas.  In this case $\hat{M}_1$ and
$\hat{M}_2$ commute, and hence can be simultaneously diagonalized.

Let the system and ancilla initially be in a state $\rho\otimes|0\rangle\langle 0|$.
Consider the unitary operation
\begin{equation}
\hat{U}(0)= \hat{M}_1\otimes \hat{Z} +  \hat{M}_2\otimes \hat{X}
,\label{uzero}
\end{equation}
where $\hat{X}$ and $\hat{Z}$ are Pauli matrices acting on the
ancilla bit. By applying $\hat{U}(0)$ to the extended
system we transform it to:
\begin{eqnarray}
\hat{U}(0)(\rho\otimes|0\rangle\langle0|)\hat{U}^{\dagger}(0)
&=& \hat{M}_1\rho \hat{M}_1\otimes|0\rangle\langle 0| \nonumber\\
&& + \hat{M}_1 \rho \hat{M}_2\otimes|0\rangle\langle 1| \nonumber\\
&& + \hat{M}_2 \rho \hat{M}_1\otimes|1\rangle\langle 0| \nonumber\\
&& + \hat{M}_2 \rho \hat{M}_2\otimes|1\rangle\langle 1|.
\end{eqnarray}
Then a projective measurement on the ancilla in the computational
basis would yield one of the possible generalized measurement
outcomes for the system. We can perform the projective
measurement on the ancilla as a sequence of weak measurements by the procedure we described
earlier. We will then prove that for this process, there exists a
corresponding sequence of generalized measurements with the same
effect acting solely on the system.  To prove this, we first show that at any
stage of the measurement process, the state of the extended system
can be transformed into the form $\rho(x)\otimes
|0\rangle\langle 0|$ by a  unitary operation
which does not depend on the state.

The net effect of the joint unitary operation $\hat{U}(0)$,
followed by the effective measurement operator on the ancilla, can
be written in a block form in the computational basis of the
ancilla:
\begin{eqnarray}
\hat{\bar{M}}(x) &\equiv& (\hat{I}\otimes\hat{P}(x))\hat{U}(0) \nonumber\\
&=& \begin{pmatrix}
\sqrt{\frac{1-\tanh(x)}{2}}\hat{M}_1&\sqrt{\frac{1-\tanh(x)}{2}}\hat{M}_2\\
\sqrt{\frac{1+\tanh(x)}{2}}\hat{M}_2&-\sqrt{\frac{1+\tanh(x)}{2}}\hat{M}_1
\end{pmatrix}.
\end{eqnarray}
If the current state $\hat{\bar{M}}(x)(\rho\otimes|0\rangle\langle 0|)\hat{\bar{M}}^\dagger$
can be transformed to
$\rho(x)\otimes|0\rangle\langle 0|$ by a unitary operator $\hat{U}(x)$ which is
independent of $\rho$, then the lower left block of $\hat{U}(x)\hat{\bar{M}}(x)$
should vanish. We look for such a unitary
operator in block form, with each block being Hermitian and
diagonal in the same basis as $\hat{M_1}$ and $\hat{M_2}$. One solution is:
\begin{equation}
\hat{U}(x)=\begin{pmatrix} \hat{A}(x)&\hat{B}(x)\\
\hat{B}(x)&-\hat{A}(x)
\end{pmatrix},
\end{equation}
where
\begin{equation}
\hat{A}(x)=\sqrt{1-\tanh(x)}\hat{M}_1(\hat{I}+\tanh(x)(\hat{M}_2^2-\hat{M}_1^2))^{-\frac{1}{2}},\label{A}
\end{equation}
\begin{equation}
\hat{B}(x)=\sqrt{1+\tanh(x)}\hat{M}_2(\hat{I}+\tanh(x)(\hat{M}_2^2-\hat{M}_1^2))^{-\frac{1}{2}}.\label{B}
\end{equation}
(Since $\hat{M}_1^2 + \hat{M}_2^2 = \hat{I}$, the operator
$(\hat{I}+\tanh(x)(\hat{M}_2^2-\hat{M}_1^2))^{-\frac{1}{2}} $
always exists.)  Note that $\hat{U}(x)$ is Hermitian, so
$\hat{U}(x)=\hat{U}^\dagger (x)$ is its own inverse, and at $x=0$
it reduces to the operator \eqref{uzero}.

After every measurement on the ancilla, depending on the value of
$x$, we apply the operation $\hat{U}(x)$. Then,
before the next measurement, we apply its inverse
$\hat{U}^{\dagger}(x)=\hat{U}(x)$. By doing this, we can think of
the procedure as a sequence of generalized measurements on the
extended system that transform it between states of the form
$\rho(x)\otimes|0\rangle\langle 0|$ (a generalized
measurement preceded by a unitary operation and followed by a
unitary operation dependent on the outcome is again a generalized
measurement). The measurement operators are now
$\hat{\tilde{M}}(x,\pm\varepsilon) \equiv
\hat{U}(x\pm\varepsilon)(\hat{I} \otimes \hat{P}(\pm\varepsilon))
\hat{U}(x)$, and have the form
\begin{equation}
\hat{\tilde{M}}(x,\pm\varepsilon)=\begin{pmatrix}\hat{M}(x,\pm\varepsilon)&
\hat{N}(x,\pm\varepsilon)\\\hat{0}&\hat{O}(x,\pm\varepsilon)
\end{pmatrix}.
\end{equation}
Here $\hat{M},\hat{N},\hat{O}$ are operators acting on the system.
Upon measurement, the state of the extended system is transformed
\begin{equation}
\rho(x)\otimes|0\rangle\langle0|\rightarrow
\frac{\hat{M}(x,\pm\varepsilon)\rho(x)\hat{M}^{\dagger}(x,\pm\varepsilon)}{p(x,\pm\varepsilon)}\otimes|0\rangle\langle0|,
\end{equation}
with probability
\begin{equation}
p(x,\pm\varepsilon) =
\tr\left\{ \hat{M}(x,\pm\varepsilon)\rho(x)\hat{M}^{\dagger}(x,\pm\varepsilon) \right\}.
\end{equation}
By imposing
$\hat{\tilde{M}}^\dagger(x,\varepsilon)\hat{\tilde{M}}(x,\varepsilon)+\hat{\tilde{M}}(x,-\varepsilon)^\dagger\hat{\tilde{M}}(x,-\varepsilon)=\hat{I}$,
we obtain that
\begin{equation}
\hat{M}^\dagger(x,\varepsilon)\hat{M}(x,\varepsilon)+\hat{M}^\dagger(x,-\varepsilon)\hat{M}(x,-\varepsilon)=\hat{I},
\end{equation}
where the operators in the last equation acts on the {\it system space alone}. Therefore, the same transformations
that the system undergoes during this procedure can be achieved by
the measurements $\hat{M}(x,\pm \varepsilon)$ {\it acting solely on the
system}. Depending on the current value of $x$, we perform the
measurement $\hat{M}(x,\pm\varepsilon)$. Due to the one-to-one
correspondence with the random walk for the projective measurement
on the ancilla, this procedure also follows a random walk with a
step size $|\varepsilon|$. It is easy to see that if the
measurements on the ancilla are weak, the corresponding
measurements on the system are also weak. Therefore we have shown
that every measurement with positive operators $\hat{M}_1$ and
$\hat{M}_2$, can be implemented as a sequence of weak
measurements.  This is the main result of this paper.  From the construction above, one can find the
explicit form of the weak measurement operators:
\begin{eqnarray}
\hat{M}(x,\varepsilon) &=& \sqrt{\frac{1-\tanh(\varepsilon)}{2}}\hat{A}(x)\hat{A}(x+\varepsilon) \nonumber\\
&& + \sqrt{\frac{1+\tanh(\varepsilon)}{2}}\hat{B}(x)\hat{B}(x+\varepsilon).\label{measpos}
\end{eqnarray}

Note that this procedure works even if the
step of the random walk is not small, since $\hat{P}(x)\hat{P}(y)\propto \hat{P}(x+y)$
for arbitrary values of $x$ and $y$. So it is not
surprising that the effective operator which gives the state of the system at the point $x$ is
\begin{eqnarray}
\hat{M}(0,x) &=& \sqrt{\frac{1-\tanh(x)}{2}}\hat{M}_1\hat{A}(x) \nonumber\\
&& + \sqrt{\frac{1+\tanh(x)}{2}}\hat{M}_2\hat{B}(x),
\end{eqnarray}
where $\hat{M}(x,y)$ is defined by \eqref{measpos}.

Finally, consider the most general type of two-outcome generalized measurement, with the only
restriction being
$\hat{M}_1^{\dagger}\hat{M}_1+\hat{M}_2^{\dagger}\hat{M}_2=I$. By
polar decomposition the measurement operators can be written
\begin{equation}
\hat{M}_{1,2}=\hat{V}_{1,2}\sqrt{\hat{M}_{1,2}^{\dagger}\hat{M}_{1,2}},
\end{equation}
where $\hat{V}_{1,2}$ are appropriate unitary operators.  One can think
of these unitaries as causing an additional disturbance to the state of the
system, in addition to the reduction due to the measurement.  The
operators $(\hat{M}_{1,2}^{\dagger}\hat{M}_{1,2})^{1/2}$ are
positive, and they form a measurement.  We could then measure
$\hat{M}_1$ and $\hat{M}_2$ by first measuring these positive operators
by a sequence of weak measurements, and then performing either
$\hat{V}_1$ or $\hat{V}_2$, depending on the outcome.

However, we can also decompose this measurement directly
into a sequence of weak measurements.  Let the weak measurement
operators for $(\hat{M}_{1,2}^{\dagger}\hat{M}_{1,2})^{1/2}$
be $\hat{M}_p(x,\pm\varepsilon)$.
Let  $\hat{V}(x)$ be any continuous unitary operator function satisfying
$\hat{V}(0)=\hat{I}$ and $\hat{V}(\pm x) \rightarrow \hat{V}_{1,2}$
as $x\rightarrow\infty$.  We then define
$\hat{M}(x,y)\equiv\hat{V}(x+y)\hat{M}_p(x,y)\hat{V}^{\dagger}(x)$.
By construction $\hat{M}(x,\pm y)$ are measurement operators.
Since $\hat{V}(x)$ is continuous, if
$y=\varepsilon$, where $\varepsilon \ll 1$, the measurements are
weak. The measurement procedure is analogous to the previous cases
and follows a random walk along the curve $\hat{M}(0,x)=\hat{V}(x)\hat{M}_p(0,x)$.

In summary, we have shown that for every two-outcome measurement
described by operators $\hat{M}_1$ and $\hat{M}_2$ acting on a
Hilbert space of dimension $d$, there exists a continuous
two-parameter family of operators $\hat{M}(x,y)$ over the same
Hilbert space with the following properties:
(1) $ \hat{M}(x,0)=\hat{I}/\sqrt{2}$,
(2) $\hat{M}(0,x) \rightarrow \hat{M}_1$ as $x\rightarrow-\infty$,
(3) $\hat{M}(0,x) \rightarrow \hat{M}_2$ as $x\rightarrow+\infty$,
(4) $ \hat{M}(x+y,z)\hat{M}(x,y)\propto\hat{M}(x,z+y)$,
(5) $\hat{M}^{\dagger}(x,y)\hat{M}(x,y) + \hat{M}^{\dagger}(x,-y)\hat{M}(x,-y) = \hat{I}$.
We have presented an explicit solution for $\hat{M}(x,y)$ in terms of
$\hat{M}_1$ and $\hat{M}_2$.
The measurement is implemented as a random walk on the curve
$\hat{M}(0,x)$ by consecutive application of the measurements
$\hat{M}(x,\pm\varepsilon)$, which depend on the current value of
the parameter $x$.  In the case where $|\varepsilon| \ll 1$, the
measurements driving the random walk are weak.  Since any
measurement can be decomposed into two-outcome
measurements, weak measurements are {\it universal}.

It is obvious from the form of the operators \eqref{measpos}
that if a measurement is {\it local}---the
measurement operators $\hat{M}_i\equiv \hat{M}_i\otimes \hat{I}$
act as the identity on all except one subsystem of a
composite system---it can be implemented as sequence of weak
\textit{local} measurements. This result may be
useful for the study of LOCC (Local Operations with Classical
Communication).

For example, a very useful concept in the theory of
entanglement is the {\it entanglement monotone} \cite{Vidal00},
a function of the state that is non-increasing on average
under local operations. For pure states the operations are
unitaries and generalized measurements. Since all unitaries can be
broken into a series of infinitesimal steps and (as we have shown)
all measurements can be decomposed into weak measurements, it
suffices to look at the behavior of a prospective monotone under
small changes in the state.  We can thus derive differential
conditions for a function to be an entanglement monotone.
This was one of the main motivations for this work.

Moreover, we can think of the set LOCC itself (or at least that
subset which preserves pure states) as being generated by
infinitesimal local operations.  This gives another way of
thinking about entanglement protocols, somewhat analogous to
studying a Lie algebra by examining the behavior of its
generators.  These topics will be the subject of a follow-up paper
\cite{monotones}.

The connection between weak measurements and
quantum trajectories is also an interesting question.
Quantum trajectories can be thought of as continuous
measurements; these generalized measurements should
therefore correspond to continuous measurements where the type
of measurement is also continuously adjusted.  This might be
experimentally feasible for some quantum optical or atomic systems,
with possible application to experiments in quantum control.
No doubt the decomposition into weak measurements will have many other
applications; it adds yet another tool to the arsenal of the
quantum information theorist.

\end{document}